\documentclass[aps,prl,reprint,groupedaddress,amsmath,amssymb,longbibliography]{revtex4-2}

\usepackage{graphicx}
\usepackage{dcolumn}
\usepackage{bm}
\usepackage{xcolor}

\begin{document}

\title{Non-adiabatic Strong Field Ionization of Atomic Hydrogen}

\author{D. Trabert$^1$}
\author{N. Anders$^1$}
\author{S. Brennecke$^2$}
\email{simon.brennecke@itp.uni-hannover.de}
\author{M. S. Sch\"offler$^1$}
\author{T. Jahnke$^1$}
\author{L. Ph. H. Schmidt$^1$}
\author{M. Kunitski$^1$}
\author{M. Lein$^2$}
\author{R. D\"orner$^1$}
\author{S. Eckart$^1$}
\email{eckart@atom.uni-frankfurt.de}

\affiliation{$^1$ Institut f\"ur Kernphysik, Goethe-Universit\"at, Max-von-Laue-Str. 1, 60438 Frankfurt am Main, Germany \\
$^2$ Institut f\"ur Theoretische Physik, Leibniz Universit\"at Hannover, Appelstr. 2, 30167 Hannover, Germany
}

\date{\today}
\begin{abstract}
We present experimental data on the non-adiabatic strong field ionization of atomic hydrogen using elliptically polarized femtosecond laser pulses at a central wavelength of 390\,nm. Our measured results are in very good agreement with a numerical solution of the time-dependent Schr\"odinger equation (TDSE). Experiment and TDSE show four above-threshold ionization (ATI) peaks in the electron's energy spectrum. The most probable emission angle (also known as 'attoclock-offset angle' or 'streaking angle') is found to increase with energy, a trend that is opposite to standard predictions based on Coulomb interaction with the ion. We show that this increase of deflection-angle can be explained by a model that includes non-adiabatic corrections of the initial momentum distribution at the tunnel exit and non-adiabatic corrections of the tunnel exit position itself.
\end{abstract}

\maketitle

Atomic hydrogen is the simplest atomic system and thus it is often used to benchmark theoretical models making it the drosophila of theories of light-matter interaction. Its key advantage is the absence of multi-electron effects and the well-defined electrostatic potential of the proton, eliminating the need for approximations. While atomic hydrogen is very frequently used in theoretical studies and textbook examples, due to the experimental challenges connected with producing hydrogen atoms in an ultra high vacuum environment and separating events from those resulting from non-dissociated $H_{2}$, only a single experiment has been reported for strong field ionization \cite{sainadh2019attosecond}. In this pioneering work, Sainadh et al. have applied the technique of angular streaking \cite{Smolarski2010,Eckle2008} to adiabatic tunnel ionization of atomic hydrogen. In angular streaking experiments, the most probable electron emission angle is analyzed and the experimentally obtained results are used to benchmark theoretical models to better understand strong field ionization \cite{torlina2015interpreting,shafir2012resolving,Bray2018,Ni2016,Camus2017,Ni2018_theo,Hofmann2019}.

In this letter, we report on a similar experiment to the one by Sainadh et al. \cite{sainadh2019attosecond} but target non-adiabatic strong field ionization \cite{Misha2005,Olga2011A,Klaiber2016,Eckart2018_Offsets} instead of the much better understood adiabatic process \cite{Ammosov1986}. Technically this is done by choosing a different intensity and wavelength regime as in \cite{sainadh2019attosecond}. For adiabatic tunnel ionization, the experimentally measured electron energy spectrum usually shows only one broad peak. We observe four above-threshold ionization (ATI) peaks \cite{voronov1966many,agostini1979p} in the electron's energy spectrum that are spaced by the photon energy, which allows for the investigation of angular streaking for each energy peak separately \cite{Goreslavski2004,WangarXiv}. To this end, atomic hydrogen is irradiated with an elliptically polarized femtosecond laser pulse at a central wavelength of 390\,nm and a peak intensity of $1.4\cdot 10^{14}$W/cm$^2$. This corresponds to a Keldysh parameter of 3, indicating that the temporal evolution of the tunnel barrier cannot be neglected under these conditions and that field-driven dynamics before and during tunneling will contribute to the ionization dynamics \cite{Misha2005}.

In our experiment, atomic hydrogen was generated using a commercially available source (H-flux, Tectra GmbH). Hydrogen gas is thermally dissociated in a tungsten capillary heated by electron bombardment. The hydrogen beam is collimated to a half opening angle of about $1^\circ$ using an aperture and reaches the laser focus after propagating in vacuum for $260\,$mm. Our optical setup is based on a commercial Ti:sapphire femtosecond laser system (100 kHz, Wyvern-500, KMLabs). We used a 200$\, \mu$m $\beta$-barium-borate (BBO) crystal to frequency double the laser pulses. The power was adjusted by a $\lambda/2$ waveplate and a subsequent thin-film polarizer. The polarization of the laser pulses was controlled using a $\lambda/4$ waveplate followed by a $\lambda/2$ waveplate. We use elliptically polarized light with an ellipticity of $\epsilon=0.85$. The duration of the laser pulses at a central wavelength of 390\,nm was $50\,$fs, and the pulses were focused by a spherical mirror ($f=60$\,mm) onto the jet of hydrogen atoms. A COLTRIMS reaction microscope \cite{ullrich2003recoil} was used to measure the three-dimensional momenta of electron and ion in coincidence. In the spectrometer, a homogeneous electric field of 27\,V/cm and a parallel homogeneous magnetic field of $14.2\,$G separated and accelerated the charged particles towards two position- and time-sensitive detectors \cite{jagutzki2002multiple}. The length of the ion (electron) spectrometer was 17\,cm (31\,cm). Each detector comprises a stack of two micro-channel plates with a diameter of 80\,mm followed by a hexagonal delay line anode.  The time-of-flight (TOF) and position-of-impact information were used to calculate the momenta of electron and ion, which were measured in coincidence. The absolute orientation of the polarization ellipse in the experiment has been determined by the most probable emission angle of $H^{+}$ ions resulting from the reaction $H_{2}\rightarrow H+H^{+}+e^{-}$ undergone by $H_{2}$ molecules \cite{Staudte2009} that were not dissociated in the atomic hydrogen source. The $H^{+}$ ions from ionization of atomic hydrogen and those from ionization of $H_{2}$ with subsequent dissociation are unambiguously distinguishable, as the latter carry a momentum of about $9$\,a.u. The average intensity in the focus was $9\cdot 10^{13}$\,W/cm$^{2}$ (taking volume averaging into account as in Ref. \cite{Wang2005}). This corresponds to a peak intensity of $1.4\cdot10^{14}$\,W/cm$^{2}$ (peak electric field of $E_\mathrm{peak}=0.046\,$a.u. at an ellipticity of $\epsilon=0.85$). The calibration of the laser intensity was done by comparing the measured electron energy distribution to numerical solutions of the TDSE. The numerical solutions of the TDSE are found in three dimensions in the length gauge for a laser pulse with a sin$^2$-envelope and a duration of 20 optical cycles. The numerical propagation uses the pseudospectral method described in Refs. \cite{TONG1997,Murakami2013,Baykusheva2017}. In order to obtain photoelectron momentum distributions, we project the final electronic wave function on the exact scattering states of the Coulomb potential. Afterwards, the distributions are averaged over four carrier-envelope phase values from 0 to 2$\pi$ and over the focal volume intensity distribution, assuming a Gaussian beam profile in the focus. We have confirmed that the observables presented in this paper do not significantly depend on the pulse length that is used in the TDSE simulation by varying the pulse length in the range between 10 and 30 optical cycles.

\begin{figure}[h]
\includegraphics[width=\columnwidth]{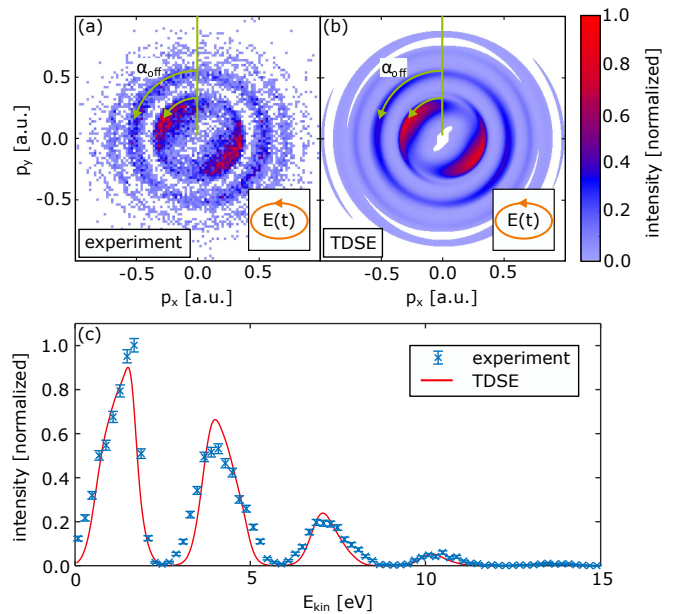}
\caption{\label{fig1} (a) Measured electron momentum distribution projected onto the laser's polarization plane for the ionization of atomic hydrogen by femtosecond laser pulses at a central wavelength of 390\,nm, an ellipticity of $\epsilon=0.85$ and a peak intensity of $1.4\cdot 10^{14}$\,W/cm$^{2}$. The light's helicity and the orientation of the polarization ellipse are indicated by the inset in the lower right corner. (b) shows a focal-averaged numerical solution of the time-dependent Schr\"odinger equation (TDSE) for the parameters that were used in (a). $\alpha_\mathrm{off}$ indicates the 'angular offset' with respect to the minor axis of the ellipse of the laser electric field and the negative vector potential in (a) and (b). The intensity in (a) [(b)] has been normalized to a maximum of 1 [0.8]. (c) shows a comparison of the electron energy distribution for the data shown in (a) and (b). The error bars show statistical errors only.}%Both, the peak positions and the relative peak heights show a high degree of agreement between experiment and TDSE.
\end{figure}

Fig. \ref{fig1}(a) shows the electron momentum distribution projected onto the laser's polarization plane ($p_{x}p_{y}$-plane). Four ATI rings are visible that are spaced by the photon energy of 3.18\,eV. The orientation of the major axis of the polarization ellipse is aligned along the $p_{x}$-direction. Fig. \ref{fig1}(b) shows the corresponding result that is obtained from a numerical solution of the time-dependent Schr\"odinger equation (TDSE). Experiment and TDSE result show excellent agreement. This is underlined by Fig. \ref{fig1}(c) which shows the corresponding electron energy distributions that are also in good agreement.

\begin{figure}[h]
\includegraphics[width=0.9 \columnwidth]{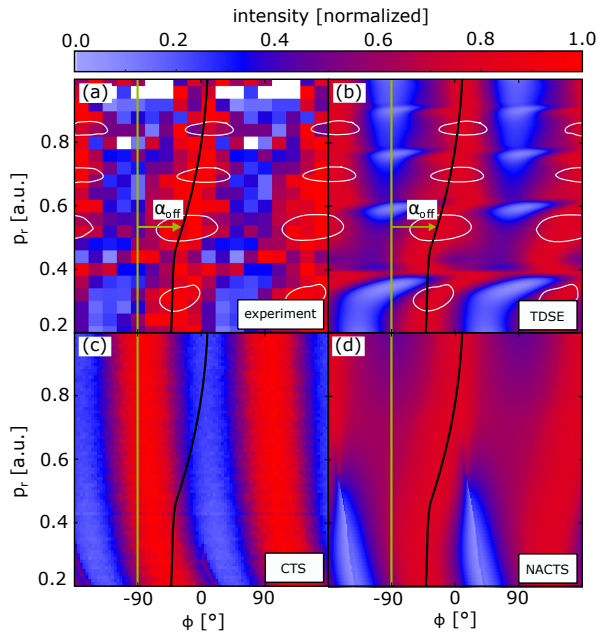}
\caption{\label{fig2} (a) [(b)] shows the electron momentum distribution from Fig. \ref{fig1}(a)  [\ref{fig1}(b)] in cylindrical coordinates  ($p_{r}=\sqrt{p_x^2+p_y^2}$ and $\phi$ is the angle of $\vec{p}_{elec}$ in the laser polarization plane) after row-wise normalization. $\alpha_\mathrm{off}$ indicates the 'angular offset', the white contours indicate the ATI peak positions. (c) shows the row-wise normalized electron momentum distribution that is obtained from the classical two-step (CTS) model with initial momenta distributed according to ADK theory, taking Coulomb interaction after tunneling into account and using the tunnel exit position from the TIPIS model in analogy to (a) and (b). (d) shows the result from our NACTS model. Within the NACTS model, the ionization probability and the initial conditions are taken from SFA. The black line guides the eye and is the same in all panels. The data in (a)-(d) has been symmetrized making use of the two-fold symmetry. Every row has been normalized independently. The intensity in (a) [(b)-(d)] has been normalized to a maximum of 1 [0.8].}
\end{figure}

In order to compare the angular offset angles $\alpha_\mathrm{off}$, the momentum distributions from Fig. \ref{fig1}(a) and (b) are shown in Fig. \ref{fig2}(a) and (b) in cylindrical coordinates. $\alpha_\mathrm{off}$ is the angle between the observed angle of maximum electron yield and the minor axis of the polarization ellipse (see Fig. 1). Row-wise normalization is performed to improve the visibility of the angular distributions for less probable $p_{r}$. Besides the good agreement of experiment and TDSE, it is evident that $\alpha_\mathrm{off}$ increases as a function of the radial momentum in the laser's polarization plane $p_{r}=\sqrt{p_x^2+p_y^2}$. What is the microscopic reason for this dependence of $\alpha_\mathrm{off}$ on $p_r$?

A common way to model strong field ionization in a time-dependent electric field $\vec{E}(t)$ is to split the ionization dynamics into two steps \cite{Gallagher1988}. First, at a given instant $t_{0}$ the electron tunnels through the time-dependent potential barrier that is formed by the ionic potential and the laser electric field. The electron is released with an initial momentum $\vec{p}_{init}$ at a position $\vec{r}_{0}$ that is anti-parallel to the electric field vector $\vec{E}(t_{0})$ at the time $t_{0}$. It is typically assumed that $\vec{p}_{init}$ is perpendicular to $\vec{E}(t_{0})$ \cite{Goreslavski2004,Shilovski2016}. In the second step, the classical forces, that act upon the electron and which are due to the laser field and Coulomb interaction with the ion, determine the electron's dynamics. For elliptically polarized light, the probability for tunneling maximizes at the two instants per laser cycle when the electric field vector points along the major axis of the polarization ellipse. Those instants in time are $t_1$ and $t_2=t_1+T/2$, where $T$ is the duration of one cycle of the light field. $\vec{p}_{init}$ is usually small and is typically modeled by a nearly Gaussian distribution \cite{Arissian2010}. Therefore, if Coulomb interaction after tunneling is neglected, this two-step model yields maximum probability for the occurrence of momenta of $\vec{p}_{elec}=-\vec{A}(t_{1})$ and $\vec{p}_{elec}=-\vec{A}(t_{2})$ giving rise to a two-fold symmetry of the electron momentum distribution in the plane of polarization. The orientation of this two-fold symmetric distribution in the plane of polarization is the basis of 'attoclock-experiments' \cite{Eckle2008,torlina2015interpreting,shafir2012resolving,Bray2018,Ni2016,Camus2017,Ni2018_theo}. The deviation of the orientation of the two-fold symmetric distribution with respect to the direction that is defined by $-\vec{A}(t_{1})$ and $-\vec{A}(t_{2})$ is referred to as 'angular offset' \cite{eckle2008attosecond}. A state-of-the-art classical model is the semiclassical two-step (SCTS) model that includes Coulomb interaction after tunneling and even allows for the modeling of interference \cite{Gallagher1988,Arbo2010,Shilovski2016,Eckart2018SubCycle}.

In the remainder of this letter, we build on the classical two-step (CTS) model \cite{Eckart2018_Offsets}, which is equivalent to the SCTS model \cite{Shilovski2016} but neglects interference. As a reference we first perform a CTS simulation in which the tunneling probability is given by the Ammosov–Delone–Krainov (ADK) theory \cite{Ammosov1986,Shilovski2016}. This is the CTS model, which is often used to model attoclock experiments \cite{Bisgaard2004,Pfeiffer2012naturalcoordinates,Ni2018_theo}. It takes Coulomb interaction after tunneling into account and uses the tunnel exit position from the TIPIS model (tunnel ionization in parabolic coordinates with induced dipole and Stark shift). In the following, we neglect the polarizability of the hydrogen atom. Inclusion of the polarizability would lead to a Stark shift of the ionization potential by about 0.13\,eV. Within the TIPIS model, the tunnel exit position is given by:
\begin{equation}
\label{TIPIS_eq}
r_\mathrm{TIPIS}(\vec{E}(t))=\frac{I_p+\sqrt{I_p^2-(4-\sqrt{8I_p})|\vec{E}(t)|}}{2|\vec{E}(t)|}
\end{equation}
Here, $I_p=0.5$\,a.u. is the ionization potential of atomic hydrogen. The tunnel exit position $r_\mathrm{TIPIS}$ depends solely on the magnitude of the instantaneous laser electric field $|\vec{E}(t)|$ which makes TIPIS an adiabatic model. The result from this CTS simulation is shown in Fig. \ref{fig2}(c). As expected, the Coulomb potential leads to a non-zero offset angle, but the effect decreases with increasing $p_r$; a trend just opposite to the findings obtained from experiment and TDSE. 

In a next step, we employ a hybrid model that combines the strength of strong-field approximation (SFA), which is the inclusion of non-adiabatic dynamics in the classically forbidden region (also referred to as tunnel), and the strength of the CTS model, which is the inclusion of Coulomb interaction after the electron is released from the classically forbidden region. We refer to this improved CTS model as non-adiabatic, classical two-step model (NACTS). The NACTS model and the CTS model are both based on classical trajectories. In contrast to the CTS model, the NACTS model uses the initial probability distribution from SFA in order to launch a swarm of classical trajectories. This implies that the momenta along the direction of the electric field at the instant of ionization can be nonzero, which is an important difference with respect to ADK theory. The same holds true for the initial position of the classical trajectories within the NACTS model, which are not simply calculated using the TIPIS model but they are also obtained from an SFA simulation \cite{Popruzhenko2008,Popruzhenko2008B}. In summary, our NACTS simulations are equivalent to the scheme presented in Ref. \cite{Brennecke2020_gouy}, except that interference of trajectories is neglected. The NACTS model results in the distribution that is shown in Fig. \ref{fig2}(d). Its exact shape is affected by the correlated initial momentum and position distribution (see supplemental material for details). The result from the NACTS model shows the same trend as observed in the experiment and the TDSE simulation. Comparing the two different classical models (Figs. \ref{fig2}(c) and \ref{fig2}(d)) it is found that the result from the NACTS model shows superior agreement with the experiment and with the result from the TDSE simulation. It should be noted, that the conclusions that are drawn from the CTS and the NACTS model   are limited by the validity of the models (see supplemental material for details). Within the CTS model there is an adiabatic tunneling step that is modeled by ADK theory. This is in contrast to SFA, which is the basis for the NACTS model. SFA also works for $\gamma>1$ and makes no assumption on the existence of tunneling in the sense of a slowly-varying potential barrier.

\begin{figure}[h]
\includegraphics[width=1\columnwidth]{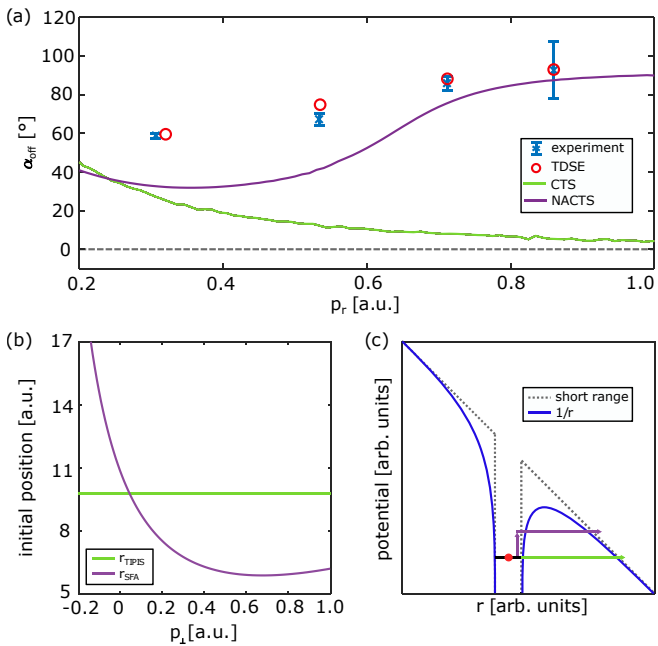}
\caption{\label{fig3} (a) The angular offset $\alpha_\mathrm{off}$ is plotted as a function of the radial momentum $p_{r}$ for the data from Fig. \ref{fig2}(a)-(d). Error bars show statistical errors only. The gray dashed line guides the eye and shows the emission angle of electrons that tunnel at the peak electric field without taking Coulomb interaction into account. The magnitude of the negative vector potential at the peak of the laser pulse is 0.33\,a.u. (b) illustrates that the tunnel exit position within the TIPIS model does not depend on the initial momentum at the tunnel exit $p_\perp$. SFA predicts decreasing values for the initial position $\vec{r}_\mathrm{SFA}$ as a function of $p_\perp$ for $0$\,a.u.$<p_\perp<0.7$\,a.u. The most probable value of $p_\perp$ from SFA is 0.15\,a.u. (c) Visualization of a potential microscopic explanation for the dependence of $\vec{r}_\mathrm{SFA}$ on $p_\perp$ (see text).}
\end{figure}

To quantitatively compare the results shown in Fig. \ref{fig2}, we extract the angular offset for each ATI peak in Fig. \ref{fig2}(a) and \ref{fig2}(b). To this end, we integrate the signal corresponding to each ATI peak separately and thereby obtain an angular distribution for each ATI peak. Then, we determine the offset angle $\alpha_\mathrm{off}$  from the second-order Fourier coefficients. For the distributions in Fig. \ref{fig2}(c) and \ref{fig2}(d), $\alpha_\mathrm{off}$ is determined for each row separately. The results are shown in Fig. \ref{fig3}(a) as a function of $p_r$. This quantitative comparison underlines the excellent agreement of the TDSE simulation with the experiment. Further, it is evident that the NACTS model qualitatively agrees with the experiment and the result from the TDSE simulation. To summarize the importance of the dependence of the initial position on the initial momentum, Fig. \ref{fig3}(b) compares $\vec{r}_\mathrm{TIPIS}$ and $\vec{r}_\mathrm{SFA}$ for tunneling at the peak electric field of an elliptically polarized light field at 390 nm with a peak electric field of 0.046\,a.u. and an ellipticity of 0.85.

The physics behind the dependence of the initial position on the initial momentum is intriguingly simple: For non-adiabatic tunneling, the energy at the tunnel exit can be higher than the ground state energy because energy can be absorbed from the light field in the classically forbidden region \cite{Misha2005}. Klaiber et al. have shown that non-adiabatic tunneling can be modeled as non-resonant multi-photon excitation and subsequent adiabatic tunneling \cite{Klaiber2015,Klaiber2016}. Within this picture, the absorption of energy from the light field leads to an inward shift of the tunnel exit position as illustrated in Fig. \ref{fig3}(c) since it effectively reduces $I_p$ in Eq. (\ref{TIPIS_eq}) by the energy that is absorbed from the light field. For almost circularly polarized light, the absorption of energy from the light field also induces a well-defined change in angular momentum. This results in an increased initial momentum at the tunnel exit \cite{Eckart2018_Offsets} and explains the overall decrease of $\vec{r}_\mathrm{SFA}$ as a function of $p_\perp$. Interestingly, for very high $p_\perp$ the energy $\frac{1}{2}p_\perp^2$ is not negligible compared to $I_p$,  which manifests as an effectively increased $I_p$, which explains the increasing values of $\vec{r}_\mathrm{SFA}$ for $p_\perp>0.7$\,a.u.

Fig. \ref{fig3}(c) visualizes that SFA does not include any long-range potential of the ion. Neglecting Coulomb effects on the under-the-barrier dynamics is a possible reason why the result from our NACTS model (see Fig. \ref{fig2}(d)) does not show complete agreement with the experiment. Further, we observe that the TDSE simulation shows a dependence of $\alpha_\mathrm{off}$ on $p_r$ even within single ATI peaks as can be seen in Fig. 2(b). Interestingly such a feature is also seen in the experiment at $p_r\approx 0.4$\,a.u. (see Fig. 2(a)). Currently we cannot explain this observation, which warrants further research. We speculate that this might be related to focal averaging, as we observe an offset angle for the lowest energy peak of only $50^{\circ}$ in the TDSE simulation without volume averaging for an intensity of $9\cdot10^{13}$\,W/cm$^{2}$.

In conclusion, we have presented a benchmark experiment on the non-adiabatic strong field ionization of atomic hydrogen. The result is in excellent agreement with our ab initio TDSE simulation. The simplicity of the structure of atomic hydrogen allows for the exclusion of multi-electron effects as well as initial states with atomic orbitals carrying angular momentum \cite{EckartNatPhys2018}. We find that the CTS model, which is typically used to interpret attoclock experiments, is not able to describe the angular offset $\alpha_\mathrm{off}$ as a function of the electron momentum in the plane of polarization $p_r$. We find that our NACTS model, a semiclassical model with non-adiabatic initial conditions and ionization probabilities from SFA, reproduces the experimental findings qualitatively. Within the NACTS model, the initial momenta and the initial positions of the wave packet at the tunnel exit are correlated, leading to a clearly improved agreement with the experiment. TDSE calculations over a wide range of Keldysh parameters (see supplemental material) suggest that the non-adiabatic effects discussed in this paper significantly alter the offset angle for $\gamma>0.5$. The inclusion of non-adiabatic effects, as in our NACTS model, can extend the validity range of two-step models to $\gamma\approx3$, which is considered outside the tunneling regime in the usual strong-field terminology. This allows for a new class of attoclock experiments that are resolved with respect to the electron’s radial momentum. We expect that the NACTS model will help to interpret future attoclock experiments and will thereby lead to a better understanding of the role of electron positions and momenta in the ionization process. The NACTS model is an important step towards probing the wave packet’s tunnel exit position with sub-\r{A}ngstrom precision \cite{SimonsneuesPaper} as proposed by Kheifets \cite{Kheifets_2020} who has referred to this approach as 'nano-ruler'. We expect that our findings will serve as a benchmark for the development of models for other atoms and small (chiral) molecules.

\section{Acknowledgments}
\normalsize
The experimental work was supported by the DFG (German Research Foundation). S.B. acknowledges funding of the German Academic Exchange Service. M.L. and S.E. acknowledge funding of the DFG through Priority Programme SPP 1840 QUTIF.

\end{document}

\title{Supplemental Material: Non-adiabatic Strong Field Ionization of Atomic Hydrogen}

\author{D. Trabert$^1$}
\author{N. Anders$^1$}
\author{S. Brennecke$^2$}
\email{simon.brennecke@itp.uni-hannover.de}
\author{M. S. Sch\"offler$^1$}
\author{T. Jahnke$^1$}
\author{L. Ph. H. Schmidt$^1$}
\author{M. Kunitski$^1$}
\author{M. Lein$^2$}
\author{R. D\"orner$^1$}
\author{S. Eckart$^1$}
\email{eckart@atom.uni-frankfurt.de}

\affiliation{$^1$ Institut f\"ur Kernphysik, Goethe-Universit\"at, Max-von-Laue-Str. 1, 60438 Frankfurt am Main, Germany \\
$^2$ Institut f\"ur Theoretische Physik, Leibniz Universit\"at Hannover, Appelstr. 2, 30167 Hannover, Germany
}

\date{\today}

\maketitle

\maketitle

\FloatBarrier
\section{I. Analytical expression for the initial position including non-adiabatic corrections}
The initial position $\vec{r}_\mathrm{SFA}$ can be determined from strong-field approximation (SFA) \cite{Popruzhenko2008,Popruzhenko2008B}. For our non-adiabatic classical two-step (NACTS) simulations, we use the exact values for $\vec{r}_\mathrm{SFA}$ as starting points for the classical trajectories. 

The initial position can also be approximated as (see Eq. (15) from Ref. \cite{Hongcheng2018}):
\begin{equation}\label{eq:1}
\vec{r}_\mathrm{SFA}(\vec{p}_\perp,\vec{E}(t),\dot{\vec{E}}(t))\approx -\frac{\vec{E}(t)}{2}\frac{p_\perp^2+2I_{p}}{|\vec{E}(t)|^2-\vec{p}_\perp\cdot\dot{\vec{E}}(t)}
\end{equation}
It is evident from Eq. (S1) that $\vec{r}_\mathrm{SFA}$ depends on the electron's initial momentum at the tunnel exit as well as the temporal evolution of the laser electric field and that $\vec{r}_\mathrm{SFA}$ and $\vec{p}_\perp$ are correlated. This is illustrated in Fig. 3(b).

\FloatBarrier
\section{II. Additional electron momentum distributions from modified semi-classical models}
Fig. S1(a) shows the result from a classical two-step (CTS) simulation that is obtained as for Fig. 2(c) but here Coulomb interaction after tunneling is neglected. This leads to an angular distribution that is independent of $p_r$ and that peaks at $-90^{\circ}$ and $+90^{\circ}$. These directions coincide with the negative vector potentials that correspond to the peaks of the magnitude of the laser electric field. Accordingly, for the data shown in Fig. S1(a) the offset angle $\alpha_\mathrm{off}$  is zero for all $p_r$. 

In Fig. S1(b) we show the result from our NACTS simulation as in Fig. 2(d) but here we use the initial position from the TIPIS model (see main text) instead of the initial position from SFA. It is evident from Fig. S1(b) that the non-adiabatic corrections of the initial position give rise to significant modifications of the electron momentum distribution. In particular, a fork-like structure is clearly visible in this distribution for radial electron momenta with $p_r>0.85$\,a.u. The fork-like structure is due to the mapping of the elliptically shaped electron momentum distribution to a cylindrical coordinate system and subsequent normalization. The branch of the fork-like structure that belongs to higher angular offsets $\alpha_\mathrm{off}$ shows a higher intensity leading to an overall shift towards higher angular offsets $\alpha_\mathrm{off}$ and improving the agreement with the experiment and the TDSE simulation. In the full NACTS model (see Fig. 2(d)), the branch that belongs to higher angular offsets $\alpha_\mathrm{off}$ shows an even higher intensity. This leads to higher angular offsets and shows the correct trend for $\alpha_\mathrm{off}$  as a function of $p_r$ that is in agreement with experiment and TDSE (see Fig. 2(a), 2(b) and 3(a)).

\begin{figure}
\includegraphics[width=\columnwidth]{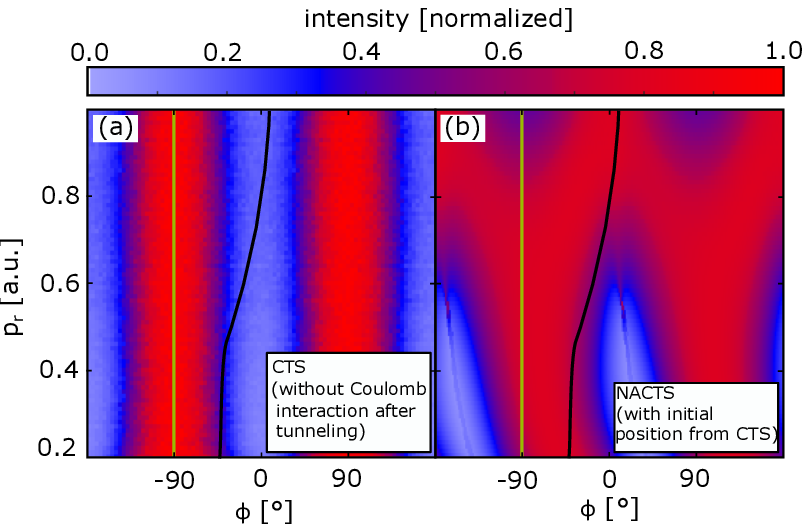}
\caption{\label{fig1}(a) shows the row-wise normalized electron momentum distribution that is obtained from the CTS model modified by neglecting Coulomb interaction after tunneling. (b) shows the electron momentum distribution from our NACTS model modified by taking the initial position from the TIPIS model. The data is presented in full analogy to Fig. 2.}%Both, the peak positions and the relative peak heights show a high degree of agreement between experiment and TDSE.
\end{figure}

\FloatBarrier
\section{III. The angular offset $\alpha_\mathrm{off}$ for different Keldysh parameters $\gamma$}
In Fig. S2 we show additional calculations for longer wavelengths and, hence, smaller Keldysh parameters compared to the parameters used in the main text. In the TDSE calculations, the angular offset $\alpha_\mathrm{off}$ increases as a function of the radial momentum $p_r$ for all studied laser parameters. This can be qualitatively reproduced using the NACTS model. For sufficiently adiabatic conditions (here for a Keldysh parameter of $\gamma=0.5$) the CTS model with initial positions from TIPIS shows the correct trend as well. Thus, we find that the increase of the angular offset $\alpha_\mathrm{off}$ as a function of $p_r$, that is observed in the main text, is universally present for a large range of laser parameters.
To explore the relevance of non-adiabaticity, the NACTS model can be modified by using the initial positions from the TIPIS model instead of using the initial positions from SFA. The results are shown in S1(b) for $\gamma=3$ and in Fig. S2 for four different values of $\gamma$. It is evident from Fig. S2 that the NACTS model that uses the initial positions from the TIPIS is in good agreement with the NACTS model for $\gamma<1$. In this parameter region, the momentum change that is due to the acceleration in the laser field (proportional to the vector potential) is large compared to the change in momentum that is due to the influence of the Coulomb interaction, that the electron experiences on its classical trajectory. Hence, the momentum distributions are only weakly influenced by Coulomb interaction. This observation agrees with the findings of the Rutherford-Keldysh-Model \cite{Bray2018} and the analytical R-Matrix theory \cite{Kaushal_PRA_2013_88}. However, for smaller wavelengths (resulting in a smaller magnitude of the vector potential and higher values of $\gamma$) the momentum change that is due to Coulomb interaction is relevant. In this case, the initial positions strongly affect the offset angle $\alpha_\mathrm{off}$. 
In conclusion, the NACTS model leads to a significant improvement in modelling the correct dependence of the angular offset $\alpha_\mathrm{off}$ as a function of the radial momentum $p_r$ as compared to the CTS model.

\begin{figure}
\includegraphics[width=\columnwidth]{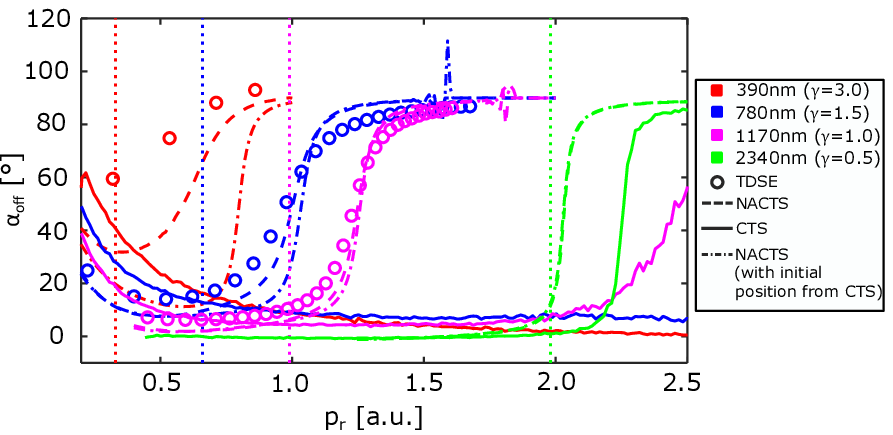}
\caption{\label{fig2} The angular offset $\alpha_\mathrm{off}$ is extracted from TDSE simulations (open circles), the NACTS model (dashed lines), the CTS model (solid lines) and the NACTS model using the initial position from the CTS model (dashed-dotted lines). Within the CTS model, the tunnel exit position from TIPIS is used (see Eq. (1)) as the initial position. The values for $\alpha_\mathrm{off}$ are plotted as a function of the radial momentum $p_r$ for different wavelengths as indicated in the legend. The corresponding Keldysh-parameters $\gamma$ are also stated. Vertical dotted lines show the absolute value of the vector potential that corresponds to ionization at the peak of the electric field. The TDSE simulations are done as for Fig. 3 but without taking volume averaging into account. We have not done a TDSE simulation for the Keldysh-parameter $\gamma=0.5$, but we expect the result to be consistent with the NACTS model. For all TDSE simulations in Fig. S2, an intensity of $9\cdot 10^{13}$W/cm$^2$ is used. CEP averaging is done as in the main manuscript.}%Both, the peak positions and the relative peak heights show a high degree of agreement between experiment and TDSE.
\end{figure}